\documentclass[letterpaper, 10 pt, conference]{ieeeconf}  

\IEEEoverridecommandlockouts                              

\let\labelindent\relax	

\usepackage{amsmath,amssymb,amsfonts}
\usepackage{mathabx}
\usepackage{algorithmic}
\usepackage{graphicx}
\usepackage{textcomp}
\usepackage[utf8]{inputenc}
\usepackage[T1]{fontenc}
\usepackage{amsmath}
\usepackage{amsfonts}
\usepackage{amssymb}
\usepackage{balance}
\usepackage{graphicx}
\usepackage[rightcaption]{sidecap}
\usepackage{import}
\usepackage{textcomp}
\usepackage[dvipsnames]{xcolor}
\usepackage{bbm}
\usepackage{slashed}
\usepackage{caption}
\usepackage{subcaption}
\usepackage{lipsum}
\usepackage{enumitem}
\usepackage{tabularx}
\usepackage[noadjust]{cite}
\usepackage{easyReview}
\usepackage{bm}
\usepackage{mathtools}
\usepackage{balance}

\newcommand{\bbR}{\mathbb{R}}

\newcommand{\calA}{\mathcal{A}}

\newcommand{\calC}{\mathcal{C}}

\newcommand{\calF}{\mathcal{F}}

\newcommand{\calH}{\mathcal{H}}

\newcommand{\calK}{\mathcal{K}}

\newcommand{\calM}{\mathcal{M}}

\newcommand{\calR}{\mathcal{R}}
\newcommand{\calS}{\mathcal{S}}

\newcommand{\calU}{\mathcal{U}}
\newcommand{\calV}{\mathcal{V}}

\newcommand{\calX}{\mathcal{X}}

\usepackage{amsthm}

\theoremstyle{definition}
\newtheorem{assumption}{Assumption}
\newtheorem{theorem}{Theorem}

\newtheorem{proposition}[theorem]{Proposition}
\newtheorem{corollary}[theorem]{Corollary} 
\newtheorem{definition}{Definition}

\newtheorem{example}{Example}

\theoremstyle{remark}
\newtheorem{remark}{Remark}

\title{\LARGE \bf
Construction of Control Barrier Functions Using Predictions with Finite Horizon
}

\author{Adrian Wiltz, Xiao Tan and Dimos V. Dimarogonas
\thanks{This work was supported by the ERC Consolidator Grant LEAFHOUND, the Horizon Europe EIC project SymAware (101070802), the Swedish Research Council, and the Knut and Alice Wallenberg Foundation.}
\thanks{The authors are with the Division of Decision and Control Systems, KTH Royal Institute of Technology, SE-100 44 Stockholm, Sweden
	{\tt\small \{wiltz,xiaotan,dimos\}@kth.se}.}%
}

\begin{document}

\maketitle
\thispagestyle{empty}
\pagestyle{empty}


\begin{abstract}
In this paper, we show that under mild controllability assumptions a time-invariant Control Barrier Function (CBF) can be constructed based on predictions with a finite horizon. As a starting point, we require only a known subset of a control-invariant set where the latter set does not need to be explicitly known. We show that, based on ideas similar to the Hamilton-Jacobi reachability analysis, the knowledge on the subset of a control-invariant set allows us to obtain a time-invariant CBF for the time-invariant dynamics under consideration. 
We also provide a thorough analysis of the properties of the constructed CBF, we characterize the impact of the prediction horizon, and comment on the practical implementation. In the end, we relate our construction approach to Model Predictive Control (MPC). With a relevant application example, we demonstrate how our method is applied.
\end{abstract}


%

\section{Introduction}

\emph{Control Barrier Functions} (CBF) have been introduced in~\cite{Wieland2007} as a control theoretic concept to render sets invariant under given dynamics. Having their origin in the optimization literature as barrier functions~\cite{Wills2004,Wright2005}, CBFs became a successfully and widely applied approach to ensure constraint satisfaction; for a survey see~\cite{Ames2019}. 
Multiple approaches to ensure constraint satisfaction via set-invariance have been considered in the literature~\cite{BlanchiniFranco2015SMiC}. However with the concept of CBFs, Control Lyapunov Functions (CLF)~\cite{Artstein1983,Sontag1989} found their control theoretic analogue for set-invariance. Both CBFs and CLFs are generally difficult to find. However, once they are found, a feedback control law is rather easily constructed.

The general task for finding a suitable CBF is as follows: Consider a dynamical system
\begin{align}
	\label{eq:system}
	\dot{x} = f(x,u)
\end{align} 
that is subject to the state constraint 
\begin{align}
	\label{eq:setH}
	x\in\calH := \lbrace x \; | \; h(x)\geq 0 \rbrace
\end{align}
where $ x\in\bbR^{n} $, $ u\in\calU\subseteq\bbR^{m} $, and $ h: \bbR^{n}\rightarrow \bbR $ is a Lipschitz continuous function. Then a Lipschitz continuous function $ b:\bbR^{n} \rightarrow \bbR $ shall be determined such that $ \calC := \lbrace x \; | \; b(x)\geq 0 \rbrace \subseteq \calH $ is a 
control-invariant subset of $ \calH $. We call such a function $ b $ a CBF. 

The task of finding a CBF is straightforward if $ f $ is locally controllable on the boundary of $ \calH $ and sufficiently large inputs $ u $ are admitted. In this case, $ h $ already constitutes a CBF. However, as soon as only weaker controllability properties hold or $ \calU $ does not allow for sufficiently large control inputs, a more sophisticated CBF construction is~required.

To this end, several approaches have been proposed. In \cite{Clark2021,Wang2022,Xu2018a}, a sum-of-squares approach is taken that determines a CBF over a basis of polynomial functions by solving an optimization problem. Such approaches are limited to polynomial dynamics and due to the complexity of the optimization problem, it is not guaranteed that a CBF is found even if it exists. 

As CBF-based feedback control laws are gradient-based \cite{Ames2019,Wieland2007}, they lead to a reactive behavior depending on the gradient of a CBF at a particular point. However, it would be advantageous if the controller already reacts before it reaches the boundary of the set $ \calC $ that shall be rendered invariant. In this way, peaks in the control signal could be avoided. Therefore, several works consider a combination of CBFs and predictive control schemes \cite{Charitidou2021,Wabersich2023}. These approaches can alleviate the problem of overly high control inputs due to their predictive nature, yet, they all assume that CBFs are readily provided. In~\cite{Gurriet2018a}, a known control-invariant set is extended by using finite horizon predictions without constructing a CBF. The first approaches to construct CBFs using a predictive strategy were proposed in~\cite{Squires2018,Breeden2021}. Here, the system dynamics controlled by a nominal feedback controller are simulated over an infinite horizon. Based on a sensitivity analysis by varying a nominal control law, \cite{Breeden2022} proposes a predictive CBF that only requires a finite time-horizon in order to ensure set-invariance. However, the time-horizon is not further specified. A different approach is taken in \cite{Choi2021}: there a combination between Hamilton-Jacobi reachability analysis and CBFs can be found which does not rely on a nominal control law. Despite the time-invariance of the state constraints under consideration, it leads to a time-dependent barrier function. This time-dependency only vanishes when the prediction horizon tends to infinity. 

In this paper, we show that under mild controllability assumptions a time-invariant CBF can be derived using a finite prediction horizon. In particular, we start from a subset of an unknown control-invariant set, and then apply finite horizon predictions to compute a CBF. Moreover, we provide a thorough analysis 
of the constructed CBF, and we relate the proposed CBF-based control strategy to MPC. 

The sequel is structured as follows. In Sec.~\ref{sec:preliminaries}, preliminaries are reviewed. In Sec.~\ref{sec:CBF construction}, we derive a CBF based on predictions with finite horizon and its properties are analyzed. In Sec.~\ref{sec:cbf and mpc}, we relate the prediction-based CBF construction to MPC. In Sec.~\ref{sec:simulation}, we present some simulations, and a conclusion is drawn in Sec.~\ref{sec:conclusion}.

\paragraph*{Notation} A continuous, strictly increasing function $ \alpha:\bbR_{\geq0}\rightarrow \bbR_{\geq0} $ with $ \alpha(0)=0 $ is called a class $ \calK $ function. A trajectory $ \bm{x}: \bbR \rightarrow \calX $ is denoted with boldface, and $ \bm{\calX}_{[t_{1},t_{2}]} $ denotes the set of all such trajectories defined on $ [t_{1},t_{2}] $. The complement of a set $ \calA\subseteq\bbR^{n} $ is denoted by $ \calA^{\text{c}} $, the Euclidean norm by $ ||\cdot|| $. The right-sided time-derivative is defined as $ \frac{d}{dt^{+}} b(x(t)) := \lim_{\varepsilon\rightarrow 0^{+}}\frac{b(x(t+\varepsilon))-b(x(t))}{\varepsilon} $. We say that a property holds almost everywhere (a.e.) if it holds everywhere except on a set of measure zero.

\section{Preliminaries}
\label{sec:preliminaries}

Throughout the paper, we consider system dynamics~\eqref{eq:system} where $ f $ is locally Lipschitz-continuous in both of its arguments to ensure the uniqueness of its solutions; forward completeness is assumed. By $ \bm{\varphi}(t;x_{0},\bm{u}) $, we denote the solution to~\eqref{eq:system} with initial state $ x(0)=x_{0} $, and input trajectory $ \bm{u}:\bbR_{\geq0}\rightarrow\calU $; the first argument $ t $ denotes the time at which $ \bm{\varphi} $ is evaluated. In the sequel, we review the most important concepts for developing our main results. 

\subsection{Control Barrier Functions}

Let $ b: \calH\rightarrow\bbR $ and define $ \calC $ as its zero-superlevel set $ \calC := \lbrace x \; | \; b(x) \geq 0 \rbrace $.
We call $ b $ a \emph{differentiable CBF} to~\eqref{eq:system} if $ b $ is differentiable and there exists a class $ \calK $ function $ \alpha $ such that for all $ x\in\calC $ it holds $ \sup_{u\in\calU} \left\lbrace \frac{\partial b}{\partial x}(x) \, f(x,u) \right\rbrace \geq -\alpha(b(x)) $.
However, confining ourselves to differentiable CBFs is limiting. Hence, we relax the differentiability of $ b $ and only require from now on that $ b $ is \emph{continuous} and \emph{piecewise differentiable}. In this case, the derivative $ \frac{d}{dt}b(x) $ cannot be expressed as $ \frac{d}{dt}b(x) = \frac{\partial b}{\partial x}(x) \, f(x,u) $ anymore because the gradient $ \frac{\partial b}{\partial x}(x) $ may not exist for $ x\in\calH $. Therefore, we replace the derivative $ \frac{d}{dt}b(x) $ by the right-sided derivative $ \frac{d}{dt^{+}}b(x) $. 
By following \cite[p.~155]{Filippov1988}, we can rewrite the derivative $ \frac{d}{dt^{+}}b(x) $ as
\begin{align}
	\label{eq:db_dt}
	\frac{d}{dt^{+}} b(x(t)) &= \frac{d}{d\sigma^{+}} b(x(t)+\sigma f(x(t),u(t)))\bigg|_{\sigma=0}
\end{align}
and state it in terms of dynamics $ f $. We can now define a broader class of CBFs as follows.
\begin{definition}[Generalized CBF]
	\label{def:generalized cbf}
	A locally Lipschitz continuous function $ b: \calH\rightarrow\bbR $ is a \emph{generalized CBF} to~\eqref{eq:system} if there exists a class $ \calK $ function $ \alpha $ such that for all~$ x\in\calC $
	\begin{align}
		\label{eq:cbf condition}
		\sup_{u\in\calU} \left\lbrace \frac{d}{d\sigma^{+}} b(x+\sigma f(x,u))\bigg|_{\sigma=0} \right\rbrace \geq -\alpha(b(x)). 
	\end{align}
\end{definition}
We call $ \calC $ \emph{control-invariant} for system~\eqref{eq:system} if there exist $ \bm{u}\in\bm{\calU}_{[0,\infty)} $ such that $ \bm{\varphi}(t;x_{0},\bm{u}) \in \calC $ for all $ t\geq0 $.
Analogously to the differentiable case, the generalized definition of CBFs also implies the control invariance of set $ \calC $.
\begin{theorem}
	\label{thm:cbf invariance}
	If $ b $ be is a generalized CBF to~\eqref{eq:system}, then $ \calC $ is control-invariant.
\end{theorem}
\begin{proof}
	As $ b $ is a generalized CBF, there exist $ u\in\calU $ due to~\eqref{eq:cbf condition} such that 
	\begin{align}
		\label{eq:cbf gradient condtion}
		\frac{d}{dt^{+}}b(x(t)) = \frac{d}{d\sigma^{+}} b(x+\sigma f(x,u))\bigg|_{\sigma=0} \geq -\alpha(b(x)).
	\end{align}
	It follows directly from the Comparison Lemma \cite[Lem.~3.4]{Khalil2002} that $ b(x(t)) \geq 0 $ for all $ t\geq 0 $ if $ b(x(0)) \geq 0 $ and control-invariance is established.
\end{proof}

\begin{remark}
	Note that differentiable CBFs are a subclass of generalized CBFs in the sense of Def.~\ref{def:generalized cbf}. When we refer to a CBF from now on, we refer to a generalized CBF.
\end{remark}

Finally, we say that a point $ x_{0}\in\bbR^{n} $ is \emph{feasible} if $ x_{0}\in\calH $, and that it is \emph{viable} if there exist $ \bm{u}\in\bm{\calU}_{[0,\infty)} $, such that $ \bm{\varphi}(t;x_{0},\bm{u})\in\calH $ for all $t\in [0,\infty) $. 

\subsection{Controllability}
\label{subsec:controllability}
A state $ x_{1} $ is time $ T $-\emph{reachable} from $ x_{0} $ for system~\eqref{eq:system} if there exists a bounded measurable input trajectory $ \bm{u}\in\bm{\calU}_{[0,T]} $, such that $ \bm{\varphi}(T;x_{0},\bm{u})=x_{1} $. The set of all such points is defined as $ \calR_{T}(x_{0}) := \lbrace x_{1} \; | \; \exists \bm{u}\in\bm{\calU}_{[0,T]}: \bm{\varphi}(
T;x_{0},\bm{u})=x_{1} \rbrace $.
We call~\eqref{eq:system} controllable on $ \calM\subseteq\bbR^{n} $ \cite{Hermann1977} if $ \bigcup_{t\in[0,\infty)} \calR_{t}(x_{0}) = \bbR^{n}, \; \forall x_{0}\in\calM $.

\section{CBF Construction}
\label{sec:CBF construction}

In order to determine if a point $ x_0 $ is viable under dynamics~\eqref{eq:system}, the trajectory $ \bm{\varphi}(t;x_0,\bm{u}) $ needs to be generally determined over an infinitely long time horizon, i.e., for all $ t\geq 0 $. By taking the system's controllability properties into account, less conservative statements can be made.

\begin{figure}[t]
	\centering
	\def\svgwidth{0.6\columnwidth}
	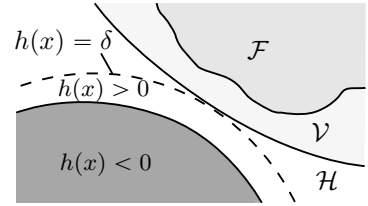
	\caption{Illustration of Ass.~\ref{ass:setF}.}
	\label{fig:setF}
	\vspace{-0.6cm}
\end{figure}

\subsection{Problem Setting}
At first, we note that if a state is sufficiently far away from the boundary of $ \calH $, we can expect it to be viable for many practically relevant systems. Therefore, it is reasonable to assume the existence of a control-invariant super-level set of~$ h $ denoted by $ \calV $ as formalized in the following assumption.

\begin{assumption}
	\label{ass:setF}
	There exists a control-invariant subset $ \calV\subset\calH $ where $ \calH:= \lbrace x \; | \; h(x)\geq 0 \rbrace $ such that $ h(x)\geq \delta $ for all $ x\in\calV $ and some $ \delta>0 $. While $ \calV $ is not required to be explicitly known, we assume that a subset $ \calF \subseteq \calV $ is known.
\end{assumption}
Ass.~\ref{ass:setF} is illustrated in Fig.~\ref{fig:setF}. Note that we do not require set $ \calF $ to be control-invariant. In many applications, a subset $ \calF\subseteq\calV $ can be more easily derived than a control-invariant set $ \calV $. Sometimes an intuitive understanding of the system dynamics can be even taken as a starting point for the derivation of $ \calF $. We illustrate this in the following example. 

\begin{figure}[t]
	\centering
	\begin{subfigure}[b]{0.4\columnwidth}
		\centering
		\def\svgwidth{1\columnwidth}
\begingroup%
  \makeatletter%
  \providecommand\color[2][]{%
    \errmessage{(Inkscape) Color is used for the text in Inkscape, but the package 'color.sty' is not loaded}%
    \renewcommand\color[2][]{}%
  }%
  \providecommand\transparent[1]{%
    \errmessage{(Inkscape) Transparency is used (non-zero) for the text in Inkscape, but the package 'transparent.sty' is not loaded}%
    \renewcommand\transparent[1]{}%
  }%
  \providecommand\rotatebox[2]{#2}%
  \newcommand*\fsize{\dimexpr\f@size pt\relax}%
  \newcommand*\lineheight[1]{\fontsize{\fsize}{#1\fsize}\selectfont}%
  \ifx\svgwidth\undefined%
    \setlength{\unitlength}{256.26495556bp}%
    \ifx\svgscale\undefined%
      \relax%
    \else%
      \setlength{\unitlength}{\unitlength * \real{\svgscale}}%
    \fi%
  \else%
    \setlength{\unitlength}{\svgwidth}%
  \fi%
  \global\let\svgwidth\undefined%
  \global\let\svgscale\undefined%
  \makeatother%
  \begin{picture}(1,0.81626357)%
    \lineheight{1}%
    \setlength\tabcolsep{0pt}%
    \put(0,0){\includegraphics[width=\unitlength,page=1]{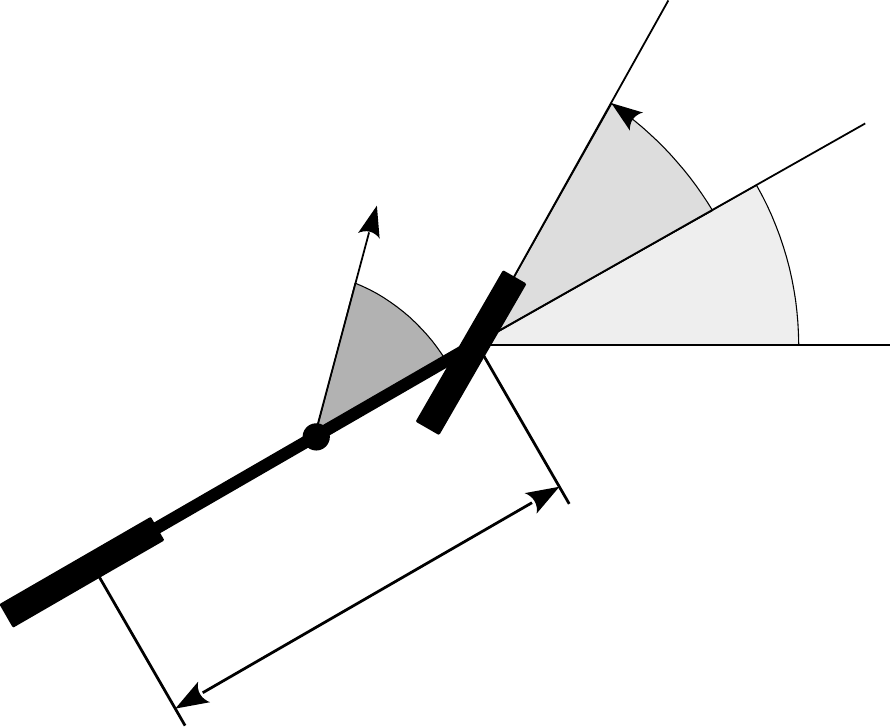}}%
    \put(0.39607649,0.40454786){\color[rgb]{0,0,0}\makebox(0,0)[lt]{\lineheight{1.25}\smash{\begin{tabular}[t]{l}$\beta$\end{tabular}}}}%
    \put(0.64354382,0.56033633){\color[rgb]{0,0,0}\makebox(0,0)[lt]{\lineheight{1.25}\smash{\begin{tabular}[t]{l}$\zeta$\end{tabular}}}}%
    \put(0.78059458,0.4530023){\color[rgb]{0,0,0}\makebox(0,0)[lt]{\lineheight{1.25}\smash{\begin{tabular}[t]{l}$\psi$\end{tabular}}}}%
    \put(0.43136955,0.06876114){\color[rgb]{0,0,0}\makebox(0,0)[lt]{\lineheight{1.25}\smash{\begin{tabular}[t]{l}$L$\end{tabular}}}}%
    \put(0.34575181,0.234386){\color[rgb]{0,0,0}\makebox(0,0)[lt]{\lineheight{1.25}\smash{\begin{tabular}[t]{l}$C$\end{tabular}}}}%
    \put(0.32587123,0.43125159){\color[rgb]{0,0,0}\makebox(0,0)[lt]{\lineheight{1.25}\smash{\begin{tabular}[t]{l}$v$\end{tabular}}}}%
    \put(0,0){\includegraphics[width=\unitlength,page=2]{bicycle.pdf}}%
  \end{picture}%
\endgroup%

		\caption{}
		\label{fig:bicycle}
	\end{subfigure}
	\hfill
	\begin{subfigure}[b]{0.55\columnwidth}
		\centering
		\def\svgwidth{1\columnwidth}
		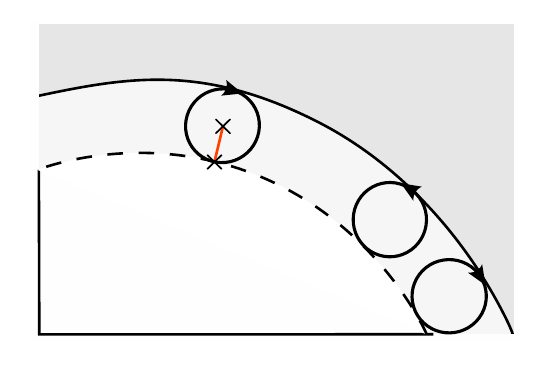
		\caption{}
		\label{fig:consrtuction safe set}
	\end{subfigure}
	\caption{Bicycle model: (a) Kinematic model; (b) construction of a set with its viable points via turning radius~$ r $.}
	\vspace{-0.6cm}
\end{figure}

\begin{example}
	\label{exmp:bicycle1}
	Consider the kinematic model of a vehicle modeled as a bicycle \cite{Wang2001} (see Fig.~\ref{fig:bicycle}) given as $ \dot{x} = v \cos(\psi + \beta(\zeta)) $, $ \dot{y} = v \sin(\psi + \beta(\zeta)) $, $ \dot{\psi} = \frac{v \cos(\beta(\zeta))\tan(\zeta)}{L} $
	where $ \beta(\zeta) = \arctan(\frac{1}{2}\tan(\zeta)) $. States $ x, y $ denote the position of the center of mass $ C $, and $ \psi $ the vehicle's orientation; inputs are velocity $ v $ and steering angle $ \zeta $. The stack vector of the system's states is denoted by $ \mathbf{x} = [x,y,\psi]^{T} $. Let the vehicle be subject to input constraints $ 0 < v_{\text{min}}\leq v\leq v_{\text{max}} $ and $ |\zeta|\leq \zeta_{\text{max}} $. From the dynamics, we directly obtain the minimal turning radius $ r=\frac{L}{\cos(\beta(\zeta_{\text{max}}))\tan(\zeta_{\text{max}})} $.
	Moreover, let the vehicle move in a plane with a circular obstacle that has radius $ R>0 $ and is centered at $ c\in\bbR^{2} $. Correspondingly, we define $ h(\mathbf{x}) = ||[x,y]^{T}-c|| - R $.
	
	Through geometric considerations as depicted in Fig.~\ref{fig:consrtuction safe set}, the set $ \calF = \lbrace \mathbf{x} \, | \, h(\mathbf{x}) \geq \delta+2r \rbrace $ is determined as a subset of the control-invariant set $ \calV $. Here we exploit the fact that a bicycle can always return to its initial state by moving on a circle with radius $ r $ and we conclude that $ h(\mathbf{x})\geq\delta $ for all $ \mathbf{x}\in\calV $. Thereby, the knowledge on the system's periodic solutions gives rise to our construction of~$ \calF $. While it is sufficient to take the vehicle's position into account for the construction of $ \calF $, the construction of $ \calV $ requires to consider orientation $ \psi $ as well. Its determination would be thereby more difficult. We highlight that $ \calV $, however, does not need to be determined and its existence is sufficient.
\end{example}

\begin{remark}
	Set $ \calF $ can be also chosen as an equilibrium point or as some periodic solution of the system. However, choosing a large $ \calF $ reduces the computational effort for determining a CBF, which is discussed later.
\end{remark}

Based on set~$ \calF $ as defined in Ass.~\ref{ass:setF}, a CBF to system~\eqref{eq:system} shall be constructed using predictions with a finite time horizon.

\subsection{Time Horizon}
\label{subsec:time horizon}
Next, we specify the minimal time $ \tau(x_{0}) $ for each $ x_{0}\in\calH\setminus\calF $ that system~\eqref{eq:system} requires to reach some state $ x_{1}\in\calF $. More precisely, we define $ \tau(x_{0}) $ for each $ x_{0}\in\calH\setminus\calF $ as
\begin{subequations}
	\label{eq:tau_x0}
	\begin{align}
		\label{seq:tau_x0 objective}
		\tau(x_{0}) &:= \min_{\tau\geq 0} \tau \\
		\label{seq:tau_x0 dynamics}
		\text{s.t.} \;\; & \dot{\bm{x}}(t) = f(\bm{x}(t),\bm{u}(t)) \quad \text{(a.e.)}, \\
		\label{seq:tau constraints}
		& \bm{x}(0) = x_{0}, \quad \bm{u}(t)\in\calU, \quad \bm{x}(\tau)\in\calF. 
	\end{align}
\end{subequations}

Note that here trajectory~$ \bm{x} $ does not need to stay in~$ \calH $ for all times. We are interested in $ \tau(x_{0}) $ for the following reasons: A trajectory that ends in $ x_{1}\in\calF $ can be feasibly continued for all times since $ \calF $ is a subset of the control-invariant set $ \calV\subset\calH $. Thereby if $ h(\bm{\varphi}(t;x_{0},\bm{u}))\geq 0 $ for all $ t\in[0,\tau(x_0)] $, we can conclude that $ x_0 $ is viable. 

In order to ensure that~\eqref{eq:tau_x0} is well-posed, that is for each $ x_{0}\in\calH\setminus\calF $ there exists a $ \tau(x_0) $ that solves~\eqref{eq:tau_x0}, the following assumption is introduced.

\begin{assumption}
	\label{ass:controllability}
	Let either of the following statements hold: 
	\begin{enumerate}[leftmargin=0.9cm]
		\item[A\ref{ass:controllability}.1] Dynamics \eqref{eq:system} are controllable on $\calF^{\text{c}}$ where $\calF^{\text{c}}$ denotes the complement of $ \calF $; or
		\item[A\ref{ass:controllability}.2] For all $ x_{0}\in\calH\setminus\calF $, there exist $ t\geq 0 $ such that $ \calR_{t}(x_{0})\cap\calF \neq \emptyset $. 
	\end{enumerate}
\end{assumption}

\begin{proposition}
	\label{prop:finite tau}
	Let Ass.~\ref{ass:controllability} hold. Then there exists for all $ x_{0}\in\calH\setminus\calF $ a time $ \tau(x_{0})\in\bbR_{>0} $ that minimizes~\eqref{eq:tau_x0}.
\end{proposition}
\begin{proof}
Starting with Ass.~\ref{ass:controllability}, it follows from the definition of controllability or time $ T $-reachability (see Sec.~\ref{subsec:controllability}) that there exists a trajectory $ \bm{\varphi}(\cdot;x_{0},\bm{u}) $ which satisfies constraints~\eqref{seq:tau_x0 dynamics}-\eqref{seq:tau constraints} for some finite $ \tau $. As $ \tau $ is lower-bounded by zero, it follows that there exists a $ \tau(x_{0}) $ that minimizes~\eqref{eq:tau_x0} for each $ x_{0}\in\calH\setminus\calF $. 
\end{proof}

\begin{remark}
	Note that (A\ref{ass:controllability}.1) is stronger than (A\ref{ass:controllability}.2) and implies the latter. (A\ref{ass:controllability}.1) can be verified using well-established criteria as the full-rank criterion for linear systems, or the Lie-rank condition for nonlinear systems \cite{Hermann1977,Sussmann1987}. (A\ref{ass:controllability}.2)~is verifiable by the construction of the reachable set. Also note that Ass.~\ref{ass:controllability} does not necessitate local controllability or even full actuation which would be strong assumptions and trivially imply the control-invariance of $ \calH $.
\end{remark}

For the sake of simplicity, we define $ \tau:= \sup_{x\in\calH\setminus\calF} \tau(x) $
and use it in the sequel instead of function $ \tau(x) $. An upper bound to $ \tau $ can often be also analytically found. 

\begin{example}
	\label{exmp:bicycle2}
	Let us revisit Example~\ref{exmp:bicycle1}. As it is well known, a bicycle can reach any point in a plane by moving on circular trajectories of radius $ r $ (by setting $ \zeta = \pm\zeta_{\text{max}} $) and straight lines ($ \zeta = 0 $). By moving with a constant velocity of at least $ v_{\text{min}} $, any position $ (x,y)\in\bbR^{2} $ can be reached in finite time. Thus~(A\ref{ass:controllability}.2) is satisfied. Next, we construct an upper-bound~$ \bar{\tau} $ for time-horizon $ \tau $ in the case of the circular obstacle from Example~\ref{exmp:bicycle1}. At first, we observe that the vehicle can reach $ \calF $ by moving on a straight line. Then the distance covered until $ \calF $ is reached is at most $ 2(R+r+\delta) $ for any starting point $ \mathbf{x}_{0}\in\calH\setminus\calF $. Alternatively, the vehicle can also move first on a semi-circle and continue thereafter on a straight line until $ \calF $ is reached. Then, the distance is upper bounded by $ \pi r + \delta + 2r $. Altogether, we obtain an upper-bound for the time-horizon as $ \bar{\tau} := \min\left\lbrace \frac{2(R+r+\delta)}{v_{\text{max}}}; \frac{(2+\pi) r + \delta}{v_{\text{max}}} \right\rbrace $.
\end{example}

\subsection{Construction of a CBF}

\begin{figure}[t]
	\centering
	\def\svgwidth{0.7\columnwidth}
	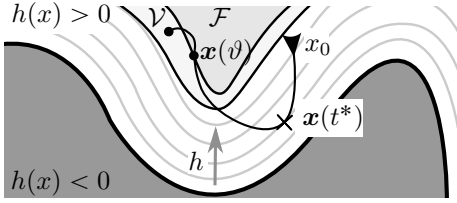
	\caption{Illustration of max-min-problem~\eqref{eq:H}: A state trajectory (black) over a time horizon $ T $ and gray contour lines denoting the values of $ h $. A control-invariant set $ \calV $ and its known subset $ \calF $ are indicated.
	}
	\label{fig:H_tilde}
	\vspace{-0.6cm}
\end{figure}

We construct a CBF based on a finite prediction horizon~$ T $. Therefore, we choose some $ T\geq \tau $, where $ \tau $ is defined as in Sec.~\ref{subsec:time horizon}, and define a function $ H_{T}: \calH \rightarrow \bbR $ as
\begin{subequations}
\label{eq:H}
	\begin{align}
		\label{seq:H max min}
		H_{T}(x_{0}) &:= \max_{\bm{u}(\cdot)} \min_{t\in[0,T]} h(\bm{x}(t)) \\
		\label{seq:H dynamics}
		\text{s.t.}\;\; &\dot{\bm{x}}(s) = f(\bm{x}(s),\bm{u}(s)) \quad (a.e.), \\
		\label{seq:H initial condition}
		&\bm{x}(0)=x_{0},\\
		\label{seq:H input constraint}
		&\bm{u}(s)\in\calU, \qquad \forall s\in[0,T]\\
		\label{seq:H terminal constraint}
		& \bm{x}(\vartheta)\in\calF, \qquad \text{for some } \vartheta\in[0,T].
	\end{align}
\end{subequations}
We denote the input trajectory $ \bm{u} $ and the times $ t $ and $ \vartheta $ that solve optimization problem~\eqref{eq:H} by $ \bm{u}^{\ast} $, $ t^{\ast} $ and $ \vartheta^{\ast} $, respectively. The computation of $ H_{T} $ is illustrated in Fig~\ref{fig:H_tilde}: The gray contour lines indicate the values of $ h $ which increase in the direction of the arrow. A state trajectory $ \bm{x}(\cdot) = \bm{\varphi}(\cdot;x_{0},\bm{u}) $ (black) starts in $ x_{0} $, evolves according to some input trajectory $ \bm{u} $ over a time-horizon $ T $, and satisfies~\eqref{seq:H terminal constraint} at time $ \vartheta\in[0,T] $. The minimization yields time $ t^{\ast} $ where the trajectory $ \bm{x}(\cdot) $ takes the smallest value on $ h $. The maximization chooses the input trajectory~$ \bm{u}^{\ast} $ such that this ``smallest value'' is as large as possible. Intuitively, $ H_{T} $ thereby provides a measure to assess how close the system state gets to the boundary of the set of allowed states~$ \calH $.
Optimization problems similar to~\eqref{eq:H} are known from Hamilton-Jacobi reachability analysis, however, without constraint~\eqref{seq:H terminal constraint}, see e.g.~\cite{Choi2021} and references therein. Thereby, our approach avoids the time-dependency of reachability value functions $ H_{T} $ by adding constraint~\eqref{seq:H terminal constraint}. 

Next, we define the zero-superlevel set of $ H_{T} $ as $ \calS_{T}:= \lbrace x \; | \; H_{T}(x)\geq 0 \rbrace $
and show that $ H_{T} $ indeed is a CBF.

\begin{theorem}
	\label{thm:cbf}
	Let Ass.~\ref{ass:setF}-\ref{ass:controllability} hold, let $ h $ be Lipschitz-continuous, and let $ T\geq \tau $. Moreover, let $ f $ be bounded on $ \calV $ in the sense that for all $ x\in\calV $ there exists a $ u\in\calU $ such that $ ||f(x,u)||\leq M $ for some constant $ M>0 $. Then $ H_{T} $ in~\eqref{eq:H} is well-defined. If $ H_{T} $ is additionally Lipschitz-continuous, then $ H_{T} $ constitutes a CBF to~\eqref{eq:system}.
\end{theorem}
\begin{proof}
	 Consider a state $ x_{0}\in\calH $. As Ass.~\ref{ass:setF} and~\ref{ass:controllability} hold, it follows from Prop.~\ref{prop:finite tau} that there exists a finite~$ \tau $ as defined in Sec.~\ref{subsec:time horizon}. This implies that for any $ T\geq\tau $ there also exists an input trajectory $ \bm{u}^{\ast}\in \bm{\calU}_{[0,T]} $ and times~$ t^{\ast} $ and $ \vartheta^{\ast} $ that solve~\eqref{eq:H}. Thereby, $ H_{T} $ is well-defined. The state trajectory induced by $ \bm{u}^{\ast} $ is denoted by $ \bm{\varphi}(\cdot;x_{0},\bm{u}^{\ast}): [0,T] \rightarrow \bbR^{n} $ and the corresponding value of $ H_{T} $ is $ H_{T}(x_{0}) = h(\bm{\varphi}(t^{\ast};x_{0},\bm{u}^{\ast})) $. In order to show that $ H_T $ is a CBF, we need to show that there exists a class~$ \calK $ function~$ \alpha $ such that $ \sup_{u\in\calU} \left\lbrace \frac{d}{dt^{+}} H_{T}(x_{0}) \right\rbrace \geq -\alpha(H_{T}(x_{0})) $ for all $ x_{0} $ with $ H_{T}(x_{0})\geq0 $. 
	 We do this in two steps. 
	 
	 \emph{Step 1:} We consider $ x_{0} $ with $ H_{T}(x_{0}) \leq \delta $. At first, we extend the input trajectory $ \bm{u}^{\ast} $ by an input trajectory $ \bm{u}_{e}\in\bm{\calU}_{[T,\infty)} $ that renders set $ \calV $ invariant. In particular, we define the extended input trajectory $ \bm{u}_{e}^{\ast} $ as 
	 \begin{align}
	 	\label{eq:cbf thm 0.5}
	 	\bm{u}_{e}^{\ast}(t) := 
	 	\begin{cases}
	 		\bm{u}^{\ast}(t) & \text{if } t\in[0,\vartheta^{\ast}]\\
	 		\bm{u}_{e}(t) & \text{if } t>\vartheta^{\ast}
	 	\end{cases}
	 \end{align}
	 where $ \bm{u}_e\in\bm{\calU}_{(\vartheta^{\ast},\infty)} $ such that $ \bm{\varphi}(t;x_{0},\bm{u}_{e}^{\ast})\in\calV $ for all $ t>\vartheta^{\ast} $. Such an input trajectory $ \bm{u}_{e}^{\ast} $ exists as $ \calV $ is control-invariant. Moreover, we observe that 
	 \begin{align}
	 	\label{eq:cbf thm 0.75}
	 	\begin{split}
	 		H_{T}(x_{0}) &= h(\bm{\varphi}(t^{\ast};x_{0},\bm{u}^{\ast})) \leq \delta \\
	 		&\stackrel{\text{Ass.~}\ref{ass:setF}}{\leq} h(\bm{\varphi}(t;x_{0},\bm{u}_{e}^{\ast})) \qquad \forall t\in[\vartheta^{\ast},\infty),
	 	\end{split}
	 \end{align}
	 and hence
	 \begin{align}
	 	\label{eq:cbf thm 1}
	 	H_{T}(x_{0}) &= h(\bm{\varphi}(t^{\ast};x_{0},\bm{u}^{\ast})) = \min_{t\in[0,T]} h(\bm{\varphi}(t;x_{0},\bm{u}^{\ast})) \nonumber\\ 
	 	&\stackrel{\eqref{eq:cbf thm 0.75}}{=} \min_{t\in[0,T')} h(\bm{\varphi}(t;x_{0},\bm{u}^{\ast}_{e}))
	 \end{align}
	 for any $ T' \geq T $. From this, it follows that
	 \begin{align}
	 	\label{eq:cbf thm 2}
	 	H_{T}(x_{0}) &= \min_{t\in[0,T)} h(\bm{\varphi}(t;x_{0},\bm{u}^{\ast}_{e})) \stackrel{\eqref{eq:cbf thm 1}}{=} \!\! \min_{t\in[0,T+t')} \!\! h(\bm{\varphi}(t;x_{0},\bm{u}^{\ast}_{e})) \nonumber\\
	 	&\leq \min_{t\in[t',T+t')}  h(\bm{\varphi}(t;x_{0},\bm{u}^{\ast}_{e})) \nonumber\\
	 	&\leq H_{T}(\bm{\varphi}(t';x_{0},\bm{u}^{\ast})) \qquad \forall t'\in[0,T]
	 \end{align}
 	 where the latter inequality holds due to the suboptimality of~$ \bm{u}_{e}^{\ast} $.
	 Thus, by again using the suboptimality of~$ \bm{u}^{\ast} $ in the first inequality, we obtain
	 \begin{align}
	 	\label{eq:cbf thm 2.5}
	 	\sup_{u\in\calU} &\left\lbrace \frac{d}{dt^{+}} H_{T}(x_{0}) \right\rbrace = \sup_{u\in\calU} \left\lbrace \frac{d}{dt^{+}} H_{T}(\bm{\varphi}(0;x_{0},u)) \right\rbrace \nonumber\\ 
	 	&\geq \frac{d}{dt^{+}} H_{T}(\bm{\varphi}(0;x_{0},\bm{u}^{\ast})) \nonumber\\ 
	 	&= \lim_{t\rightarrow 0^{+}} \frac{H_{T}(\bm{\varphi}(t;x_{0},\bm{u}^{\ast})) - H_{T}(\bm{\varphi}(0;x_{0},\bm{u}^{\ast}))}{t} \nonumber\\
	 	&= \lim_{t\rightarrow 0^{+}} \frac{H_{T}(\bm{\varphi}(t;x_{0},\bm{u}^{\ast})) - H_{T}(x_{0})}{t} \stackrel{\eqref{eq:cbf thm 2}}{\geq} 0.
	 \end{align}
	 Hence, there exists a class $ \calK $ function~$ \alpha $ such that $ \sup_{u\in\calU}\left\lbrace \frac{d}{dt^{+}} H_{T}(x_{0}) \right\rbrace \geq -\alpha(H_{T}(x_{0})) $ for all~$ x_{0} $ with $ 0\leq H_{T}(x_{0}) \leq \delta $.  
	 Intuitively,~\eqref{eq:cbf thm 2.5} implies that in the neighborhood of any point~$ x_{0} $ with $ H_{T}(x_{0})\leq\delta $ there exists a state trajectory, namely $ \bm{\varphi}(\cdot;x_{0},\bm{u}^{\ast}) $, along which the value of $ H_{T} $ is monotonously increasing. 
	 
	 \emph{Step 2:} Next we consider $ x_{0} $ with $ H_{T}(x_{0}) > \delta $, which implies $ x_{0}\in\calV $. Thus, $ ||f(x,u)||<M $ for all $ x $ in some $\epsilon$-neighborhood of $ x_{0} $ and some $ u\in\calU $. Then, as $ x(t) = x_{0} + \int_{0}^{t} f(x(s),u(s)) ds $, it follows for a sufficiently small $ t' $ 
 	 \begin{align}
 	 	\label{eq:cbf thm 3}
 	 	\begin{split}
 	 		||x_{0} &- x(t')|| = \bigg|\bigg|\int_{0}^{t'} f(x(s),u(s)) ds\bigg|\bigg| \\
 	 		&\hspace{-0.1cm}\leq \int_{0}^{t'} ||f(x(s),u(s))|| ds \leq \int_{0}^{t'} M ds = M\, t'.
 	 	\end{split}
 	 \end{align}    
 	 Let $ \bm{u}'\in\bm{\calU}_{[0,t']} $ be some input trajectory. Since $ H_{T} $ is assumed to be Lipschitz-continuous -- let $ L $ be its Lipschitz constant --, we can lower-bound $ H_{T}(\bm{\varphi}(t';x_{0},\bm{u}')) $ as
 	\begin{align}
 		\label{eq:cbf thm 5}
 		\begin{split}
 			H_{T}(\bm{\varphi}(t';x_{0},\bm{u}')) &\geq H_{T}(x_{0}) \!-\! L \, ||x_{0} \!-\! \bm{\varphi}(t';x_{0},\bm{u}')|| \\
 			&\stackrel{\eqref{eq:cbf thm 3}}{\geq} H_{T}(x_{0}) - L \, M \, t'.
 		\end{split}
 	\end{align}
 	 Now we can derive analogously to~\eqref{eq:cbf thm 2.5} that 
 	 \begin{align}
 	 	\label{eq:cbf thm 6}
 	 	\sup_{u\in\calU} &\left\lbrace \frac{d}{dt^{+}} H_{T}(x_{0}) \right\rbrace \geq \frac{d}{dt^{+}} H_{T}(\bm{\varphi}(0;x_{0},\bm{u}')) \nonumber\\ 
 	 	&= \lim_{t\rightarrow 0^{+}} \frac{H_{T}(\bm{\varphi}(t;x_{0},\bm{u}')) - H_{T}(x_0)}{t} \nonumber\\
 	 	&\stackrel{\eqref{eq:cbf thm 5}}{\geq} - L M \geq -\alpha(H_{T}(x_{0}))
 	 \end{align}
  	 for all $ x_{0} $ with $ H_{T}(x_{0}) \!>\! \delta $ where $ \alpha $ is some class~$ \calK $ function.
	 
	 Altogether, we have shown that $ \sup_{u\in\calU} \left\lbrace \frac{d}{dt^{+}} H_{T}(x(t)) \right\rbrace \geq -\alpha(H_{T}(x)) $ for all $ x_{0} $ with $ H_{T}(x_{0})\geq0 $. Together with the assumption on the local Lipschitz continuity of $ H_{T} $, the proof is completed.
\end{proof}

\begin{remark}
	After the computation of $ H_{T} $, its Lipschitz-continuity can be easily checked numerically. The boundedness of $ f $ naturally arises when considering compact sets.
\end{remark}

The CBF $ H_{T} $ gives also rise to further CBFs.

\begin{corollary}
	Let the same premises hold as in Thm.~\ref{thm:cbf}. Then $ H_{T}^{\delta'}(x) := H_{T}(x)-\delta' $ is a CBF for any $ \delta' \in [0,\delta) $.
\end{corollary}
\begin{proof}
	This follows directly from the proof of Thm.~\ref{thm:cbf} as~\eqref{eq:cbf thm 2.5} and~\eqref{eq:cbf thm 6} stay unchanged.
\end{proof}


Next we show that the CBF $ H_{T} $ renders a subset of $ \calH $, namely~$ \calS_{T} $, control-invariant.

\begin{proposition}
	\label{prop:control invariance}
	$ \calS_{T} $ is control-invariant for any $ T\geq\tau $ and $ \calS_{T}\subseteq\calH $. 
\end{proposition}
\begin{proof}
	Since $ H_{T} $ is a CBF according to Thm.~\ref{thm:cbf}, control-invariance of $ \calS_{T} $ follows directly from Thm.~\ref{thm:cbf invariance}. Furthermore, 
	\begin{align}
		\label{eq:cor control invariance_aux1}
		\begin{split}
			H_{T}(x_{0}) &\stackrel{\eqref{seq:H max min}}{=} \max_{\bm{u}(\cdot)} \min_{t\in[0,T]} h(\bm{\varphi}(t;x_{0},\bm{u})) \\
			&\; \leq \max_{\bm{u}(\cdot)} h(\bm{\varphi}(0;x_{0},\bm{u})) = h(x_{0}).
		\end{split}
	\end{align}
	If $ H_{T}(x_{0}) \geq 0 $, then $ h(x_{0}) \geq 0 $ and $ \calS_{T}\subseteq\calH $ follows. {\color{white}.}
\end{proof}

Generally by increasing $ T $, $ \calS_{T} $ can be enlarged. This is formally stated as follows.

\begin{proposition}
	Let the same premises hold as in Thm.~\ref{thm:cbf}. Then $ \calS_{T_{1}} \subseteq \calS_{T_{2}} $ for $ T_{1}\leq T_{2} $. If additionally $ H_{T_{1}}(x) < \delta $, then $ H_{T_{1}}(x) \leq H_{T_{2}}(x) $. 
\end{proposition}
\begin{proof}
	Consider any $ x_{0}\in\calS_{T_{1}} $ and a trajectory $ \bm{\varphi}(t;x_{0},\bm{u}_{T_{1}}^{\ast}) $ starting in $ x_{0} $ which is defined for $ t\in[0,T_{1}] $ and where~$ \bm{u}_{T_{1}}^{\ast} $, $ t^{\ast}_{T_{1}} $ and $ \vartheta^{\ast}_{T_{1}} $ denote the input trajectory and the times that solve~\eqref{eq:H} for time horizon~$ T_{1} $. Furthermore, we define an extended input trajectory analogously to \eqref{eq:cbf thm 0.5} as $ \bm{u}_{T_{1},e}^{\ast} := \left\lbrace\begin{smallmatrix*}[l]
		\bm{u}_{T_{1}}^{\ast}(t) & \text{if } t\in[0,\vartheta^{\ast}_{T_{1}}]\\
		\bm{u}_{e}(t) & \text{if } t>\vartheta^{\ast}_{T_{1}}
	\end{smallmatrix*} \right. $
	where $ \bm{u}_{e}(t)\in\bm{\calU}_{(\vartheta^{\ast}_{T_{1}},\infty)} $ such that $ \bm{\varphi}(t;x_{0},\bm{u}_{T_{1},e}^{\ast})\in\calV $ for all $ t>\vartheta^{\ast}_{T_{1}} $. Based on this, it follows with $ H_{T_{1}}(x_{0}) < \delta $ that 
	\begin{align}
		\label{eq:prop safe set time horizon 1}
		H_{T_{1}}(x_{0}) &= \!\!\! \min_{t\in[0,T_{1}]}\!\!\! h(\bm{\varphi}(t;x_{0},\bm{u}_{T_{1}}^{\ast})) \!\stackrel{\eqref{eq:cbf thm 1}}{=}\!\!\! \min_{t\in[0,T_{2}]}\!\!\! h(\bm{\varphi}(t;x_{0},\bm{u}_{T_{1},e}^{\ast})) \nonumber\\
		&\leq \!\min_{t\in[0,T_{2}]}\! h(\bm{\varphi}(t;x_{0},\bm{u}_{T_{2}}^{\ast})) = H_{T_{2}}(x_{0})
	\end{align}
	where the last inequality follows from the suboptimality of $ \bm{u}_{T_{1},e}^{\ast} $, and where $ \bm{u}_{T_{2}}^{\ast} $ denotes the input trajectory that solves~\eqref{eq:H} for time horizon~$ T_{2} $. For $ x_{0} $ with $ H_{T_{1}}(x_{0}) \geq \delta $, we have that 
	\begin{align*}
		H_{T_{2}}(x_{0})\! &= \!\!\min_{t\in[0,T_{2}]}\!\! h(\bm{\varphi}(t;x_{0},\bm{u}_{T_{2}}^{\ast})) \!\geq  \!\!\min_{t\in[0,T_{2}]}\!\! h(\bm{\varphi}(t;x_{0},\bm{u}_{T_{1},e}^{\ast})) \\
		&= \min \left\lbrace H_{T_{1}}(x_{0}), \min_{t\in[T_{1},T_{2}]} h(\bm{\varphi}(t;x_{0},\bm{u}_{T_{1},e}^{\ast})) \right\rbrace \geq \delta
	\end{align*}
	where the last inequality holds since $ \bm{\varphi}(t;x_{0},\bm{u}_{T_{1},e}^{\ast})\in\calV $ for all $ t\in[T_{1},T_{2}] $ and $ h(x)\geq \delta $ for all $ x\in\calV $. Thus, $ \left(H_{T_{1}}(x) \geq 0\right) \Rightarrow \left(H_{T_{2}}(x) \geq 0\right) $ which implies $ \calS_{T_{1}} \subseteq \calS_{T_{2}} $ for $ T_{1} \leq T_{2} $. The second part of the proposition has been shown in~\eqref{eq:prop safe set time horizon 1}.
\end{proof}

\subsection{Construction of a CBF using a Terminal Constraint}
By employing stronger assumptions on $ \calF $, the optimization problem~\eqref{eq:H} can be reformulated with a terminal constraint.

\begin{assumption}
	\label{ass:control invariance}
	$ \calF $ is control-invariant under dynamics~\eqref{eq:system}. 
\end{assumption}

Now, we can reformulate~\eqref{eq:H} with a terminal constraint based on set $ \calF $ as 
\begin{subequations}
	\label{eq:H simp}
	\begin{align}
		\label{seq:H max min simp}
		H_{T}(x_{0}) &:= \max_{\bm{u}(\cdot)} \min_{t\in[0,T]} h(\bm{x}(t)) \\
		\label{seq:H dynamics simp}
		\text{s.t.}\;\; &\text{\eqref{seq:H dynamics}-\eqref{seq:H input constraint} hold,} \\
		\label{seq:H terminal constraint simp}
		& \bm{x}(T)\in\calF.
	\end{align}
\end{subequations}
Compared to~\eqref{eq:H}, constraint~\eqref{seq:H terminal constraint simp} is the only difference. 
Note that Ass.~\ref{ass:control invariance} implies Ass.~\ref{ass:controllability} which leads us to the following result based on Thm.~\ref{thm:cbf}.

\begin{corollary}
	\label{thm:cbf simp}
	Let Ass.~\ref{ass:setF} and~\ref{ass:control invariance} hold, let $ h $ be Lipschitz-continuous, and let $ T\geq \tau $. Moreover, let $ f $ be bounded on $ \calV $ in the sense that for all $ x\in\calV $ there exists a $ u\in\calU $ such that $ ||f(x,u)||\leq M $ for some constant $ M>0 $. Then $ H_{T} $ in~\eqref{eq:H simp} is well-defined. If $ H_{T} $ is additionally Lipschitz-continuous, then $ H_{T} $ constitutes a CBF to~\eqref{eq:system}.
\end{corollary}

\begin{figure}[t]
	\centering
	\def\svgwidth{0.55\columnwidth}
	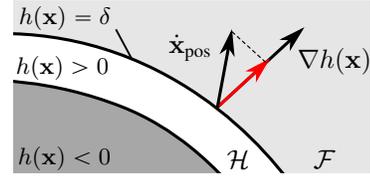
	\caption{Construction of a control-invariant set $ \calF $.}
	\label{fig:invariantset}
		\vspace{-0.5cm}
\end{figure}

\begin{example}
	\label{exmp:bicycle3} We reconsider Example~\ref{exmp:bicycle1} and construct a set~$ \calF $ that satisfies Ass.~\ref{ass:control invariance}. Let us denote the vehicle's position as $ \mathbf{x}_{\text{pos}} = [x,y]^{T} $. Then, a control-invariant set is given by $ \calF = \lbrace \mathbf{x} \, | \, h(\mathbf{x})\geq\delta, \; \nabla h(\mathbf{x})\,\dot{\mathbf{x}}_{\text{pos}}\geq 0 \rbrace $, see Fig.~\ref{fig:invariantset}. The scalar product $ \nabla h(\mathbf{x})\,\dot{\mathbf{x}}_{\text{pos}} $ is indicated by the red arrow. Intuitively, set~$ \calF $ contains those states where the vehicle is moving into its interior. Clearly, the here constructed~$ \calF $ is smaller compared to that in Example~\ref{exmp:bicycle1}. Similarly to Example~\ref{exmp:bicycle2}, an upper bound of $ \tau $ is obtained as $ \bar{\tau} = \min\left\lbrace \frac{2(R+\delta)}{v_{\text{max}}}; \frac{\pi r + \delta}{v_{\text{max}}} \right\rbrace $.
\end{example}

\subsection{Practical Computation of a CBF}

A CBF $ H_{T} $ can be computed by sampling the state space and evaluating either of the optimization problems~\eqref{eq:H} or~\eqref{eq:H simp} in each sampled state. For implementation, we rewrite the optimization as a min-max problem and use a p-norm to approximate the inner maximization problem. In this way, a stacked optimization is avoided.

Based on the computed CBF, a gradient can be numerically determined which we denote by $ \nabla_{x}H_{T} $. Using interpolation, $ H_{T}(x) $ and $ \nabla_{x}H_{T}(x) $ can be approximated for any state $ x $. Then, a safety filter for some given nominal control input $ u_{\text{nom}} $ is 
\begin{subequations}
	\label{eq:safety filter}
	\begin{align}
		\label{seq:safety filter objective}
		&u = \underset{u\in\calU}{\text{argmin}} \; (u-u_{\text{nom}})^{T}(u-u_{\text{nom}}) \\
		\label{seq:safety filter constraint}
		&\text{s.t. } \nabla_{x}H_{T}(x)\, f(x,u) \geq -\alpha(H_{T}(x))
	\end{align}
\end{subequations}
The control input $ u $ is applied to the system and ensures that the system state does not leave $ \calH $. We note that~\eqref{eq:safety filter} becomes a quadratic program if system~\eqref{eq:system} is input affine. 

The curse of dimensionality is here alleviated by the fact that the CBF only needs to be computed in some neighborhood of the boundary of $ \calH $. In particular, if $ \alpha(H_{T}(x)) $ is chosen sufficiently large for $ H_{T}(x)\geq\delta $, the optimization problem~\eqref{eq:safety filter} is minimized by $ u_{\text{nom}} $. In this way, it is sufficient to compute the CBF $ H_{T}(x) $ only for those sampled states $ x\in\calH\setminus\calF $ where $ H_{T}(x) < \delta $. Based on the systems local controllability properties, the region where $ H_T(x)< \delta $ can be determined as demonstrated in the example below. 

\begin{figure}[t]
	\centering
	\begin{subfigure}[b]{0.5\columnwidth}
		\centering
		\def\svgwidth{0.9\columnwidth}
		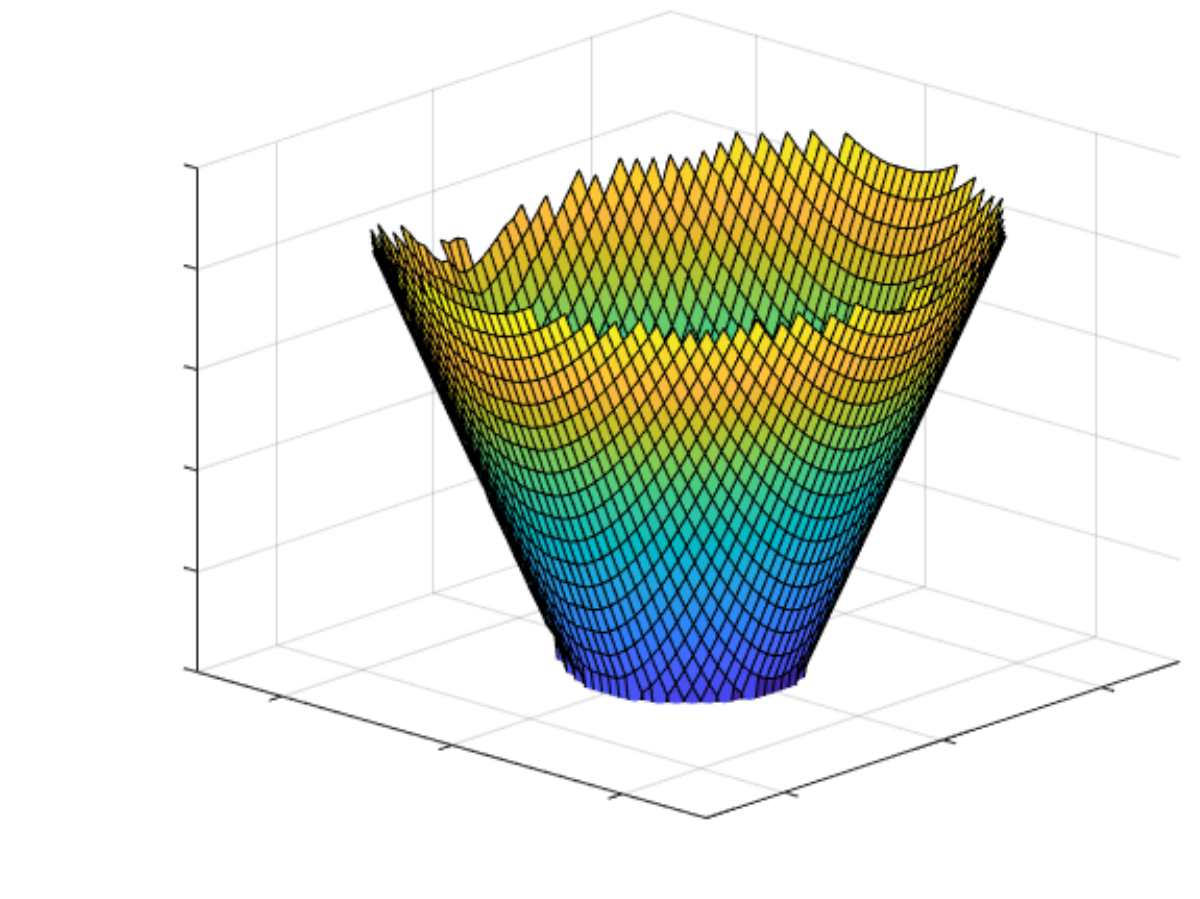
		\caption{CBF exemplarily for $ \psi=0^{\circ}. $}
		\label{fig:cbf one circ}
	\end{subfigure}
	\hfill
	\begin{subfigure}[b]{0.45\columnwidth}
		\centering
		\def\svgwidth{0.9\columnwidth}
		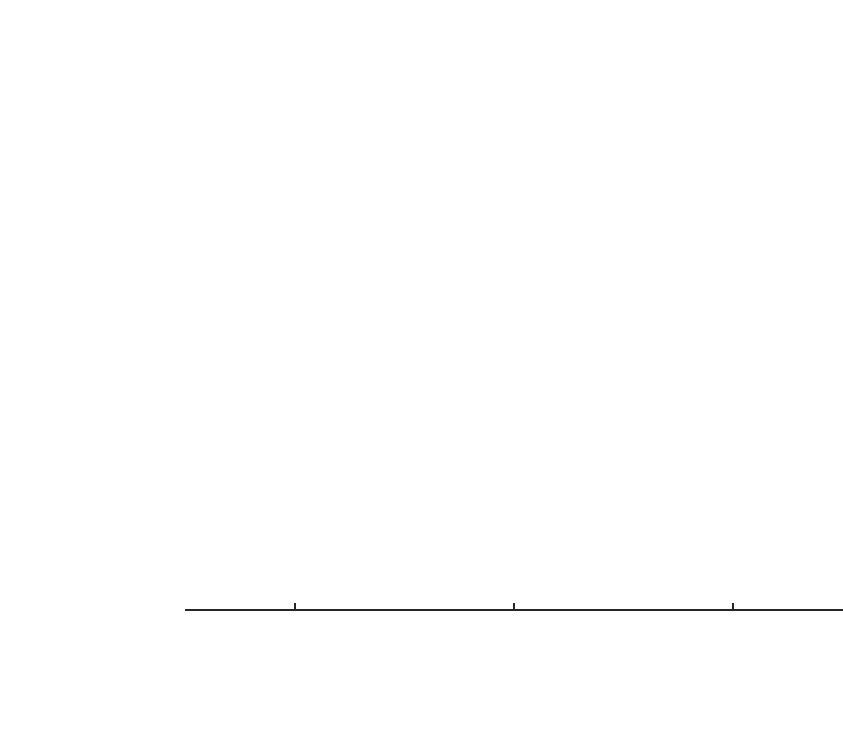
		\caption{Simulation results.}
		\label{fig:one circ sim}
	\end{subfigure}
	\caption{Scenario with one circular obstacle.}
	\vspace{-0.5cm}
\end{figure}

\begin{example}
	\label{exmp:bicycle4}
	Let us revisit Example~\ref{exmp:bicycle1} and compute a CBF for the circular obstacle and the vehicle with bicycle dynamics. Therefore, let $ R=5 $, $ L=1 $, $ v_{\text{min}} = 1 $, $ v_{\text{max}} = 5 $, $ \zeta_{\text{max}} = \frac{1}{9}\pi $, $ c = [0,0]^{T} $. Then, we obtain the minimal turning radius as $ r=2.79 $. For computing the CBF, we use~\eqref{eq:H simp} with set~$ \calF $ as constructed in Example~\ref{exmp:bicycle3}, and by choosing $ \delta=r $ we obtain $ \bar{\tau} = 4 $. By reconsidering Fig.~\ref{fig:consrtuction safe set}, we derive that $ H_{T}(x)\geq \delta $ if $ h(x)\geq 2r+\delta = 3r $. Thus, we only need to compute $ H_{T}(x) $ for all $ x\in\calH $ with $ h(x)<3r $. The result for $ T=6 $ is depicted in Fig.~\ref{fig:cbf one circ}. As it can be seen, the computed CBF is Lipschitz-continuous. We use CasADi~\cite{Andersson2019} to implement the optimization problem~\eqref{eq:H simp}.
\end{example}


\noindent In the light of this discussion, we briefly relate~\eqref{eq:H simp} to MPC.

\section{Relation of the Prediction-Based CBF Construction and MPC}
\label{sec:cbf and mpc}

In the previous section, we constructed a CBF on the grounds of a finite prediction horizon $ T $. Therefore, it seems natural to relate the CBF construction approach to a MPC formulation for ensuring set-invariance. To this end, consider the following MPC-problem at some time $ t_{k} $
\begin{subequations}
	\label{eq:MPC1}
	\begin{align}
		\label{seq:MPC max min}
		&\hspace{-0.9cm}\max_{\bm{u}(\cdot;t_{k})} \min_{t\in[0,T]} h(\bm{x}(t;t_{k})) \\
		\label{seq:MPC dynamics}
		\text{s.t.}\;\;&\dot{\bm{x}}(s;t_{k}) = f(\bm{x}(s;t_{k}),\bm{u}(s)) \quad (a.e.), \\
		\label{seq:MPC initial condition}
		&\bm{x}(0;t_{k})=\bm{x}(t_{k}),\\
		\label{seq:MPC input constraint}
		&\bm{u}(s;t_{k})\in\calU, \qquad \forall s\in[0,T] \\
		\label{seq:MPC terminal constraint}
		& \bm{x}(\vartheta;t_{k})\in\calF, \qquad \text{for some } \vartheta\in[0,T].
	\end{align}
\end{subequations}
where $ \bm{x}(\cdot;t_{k}) $, $ \bm{u}(\cdot;t_{k}) $ denote the state and input trajectories predicted at time $ t_{k} $, $ T $ is some time horizon satisfying $ T\geq\tau$, and $ \bm{x}(t_{k}) $ is the state of system~\eqref{eq:system} at time $ t_{k} $. 

The input trajectory $ \bm{u}^{\ast}(\cdot;t_{k}) $ that minimizes~\eqref{eq:MPC1} is applied to system~\eqref{eq:system} on a time-interval of length $ \Delta t \leq T $, i.e., $ \bm{u}(t)=\bm{u}^{\ast}(t;t_{k}) $ for all $ t\in[0,\Delta t) $. 
At the next time-step $ t_{k+1} := t_{k}+\Delta t $, \eqref{eq:MPC1} is recomputed with the new system state $ \bm{x}(t_{k+1}) $. If the system state is initialized as $ \bm{x}(0)\in\calS_{T} $, then we can show that $ \bm{x}(t)\in\calS_{T} $ for all $ t\geq0 $.

\begin{theorem}
	\label{thm:MPC recursive feasibility}
	Let the same premises hold as in Thm.~\ref{thm:cbf}, and let $ \calS_{T} = \lbrace x \; | \; H_{T}(x)\geq 0 \rbrace $ where $ H_{T} $ is as defined in~\eqref{eq:H}. 
	If $ \bm{x}(0)\in\calS_{T} $, then an initially feasible solution to~\eqref{eq:MPC1} exists and~\eqref{eq:MPC1} is recursively feasible for $ t_{k}\geq0 $. Moreover, $ \bm{x}(t) \in \calS_{T} $ for all $ t\geq 0 $ .
\end{theorem}
\begin{proof}
	According to Thm.~\ref{thm:cbf}, \eqref{eq:MPC1} is well-defined for all $ \bm{x}(t_{k})\in\calH $. By the premises of Thm.~\ref{thm:cbf} and by Prop.~\ref{prop:control invariance}, we have $ \bm{x}(0)\in\calS_{T}\subseteq\calH $ and the initial feasibility of~\eqref{eq:MPC1} follows. We take this as the induction base, and prove the rest of the theorem by induction. In the remainder of the proof, $ \bm{u}^{\ast}(\cdot;t_{k}) $, $ t^{\ast} $ and $ \vartheta^{\ast} $ denote the optimal input trajectory and times, respectively, that solve~\eqref{eq:MPC1} at time $ t_{k} $. The corresponding optimal state trajectory is $ \bm{x}^{\ast}(\cdot;t_{k}):=\bm{\varphi}(\cdot;\bm{x}(t_{k}),\bm{u}^{\ast}(\cdot;t_{k})) $.
	
	\emph{Induction step:} Next, we show that if $ \bm{x}(t_{k})\in\calS_{T} $, then $ \bm{x}(t_{k}+t')\in\calS_{T} $ for all $ t'\in[0,\Delta t] $. Therefore, we note that $ \bm{x}(t_{k}+t) = \bm{\varphi}(t; \bm{x}(t_{k}), \bm{u}^{\ast}(\cdot;t_{k})) = \bm{x}^{\ast}(t;t_{k}) $ and that~\eqref{seq:MPC max min} subject to (\ref{eq:MPC1}b-e) is equal to $ H_{T}(\bm{x}(t_{k})) $ as defined in~\eqref{eq:H}. 
	Next, we note that for any $ t'\in[0,T] $ it holds
    \begin{align}
		\label{eq:MPC thm 0}
		&H_{T}(\bm{\varphi}(t';\bm{x}(t_k),\bm{u}^{\ast})) \geq \min_{t\in[t',T+t']} h(\bm{\varphi}(t;\bm{x}(t_k),\bm{u}_{e}^{\ast})) \nonumber\\
		&\quad\!\!\geq \min_{t\in[0,T+t']} h(\bm{\varphi}(t;\bm{x}(t_k),\bm{u}_{e}^{\ast}))) \nonumber\\ 
		&\quad\!\!= \min \left\lbrace H_{T}(\bm{x}(t_k)), \!\min_{t\in[T,T+t']} \! h(\bm{\varphi}(t;\bm{x}(t_k),\bm{u}_{e}^{\ast})) \right\rbrace,
	\end{align}
	where the first inequality holds due to the suboptimality of $ \bm{u}_{e}^{\ast} $ which is an extended input trajectory defined as $ \bm{u}_{e}^{\ast}(t;t_{k}) := \left\lbrace \begin{smallmatrix*}[l]
		\bm{u}^{\ast}(t;t_{k}) & \text{if } t\in[0,\vartheta^{\ast}]\\
		\bm{u}_{e}(t;t_{k}) & \text{if } t>\vartheta^{\ast}
	\end{smallmatrix*} \right. $
	with $ \bm{u}_{e}\in\bm{\calU}_{(\vartheta^{\ast},\infty)} $ such that $ \bm{\varphi}(t;\bm{x}(t_{k}),\bm{u}_{e})\in\calV $ for all $ t> \vartheta^{\ast} $. The definition of $ \bm{u}_{e} $ implies $ h(\bm{\varphi}(t;\bm{x}(t_{k}),\bm{u}_{e}))\geq\delta $ for all $ t\geq \vartheta^{\ast} $. Now, we can derive
	\begin{align}
		\label{eq:MPC thm 1}
		&H_{T}(\bm{x}(t_{k+1})) = H_{T}(\bm{\varphi}(\Delta t; \bm{x}(t_{k}), \bm{u}^{\ast}(\cdot;t_{k}))) \nonumber\\
		&\!\stackrel{\eqref{eq:cbf thm 2},\eqref{eq:MPC thm 0}}{\geq} \!\!\!\!\!\min \!\left\lbrace\! H_{T}(\bm{x}(t_{k})), \!\!\!\min_{t\in[T,T+t']}\!\!\!\!\!  h(\bm{\varphi}(t;\bm{x}(t_{k}),\bm{u}_{e}^{\ast}(\cdot;t_{k})) \!\right\rbrace\! 
	\end{align}
	for all $ t'\in[0,\Delta t] $. Then, since $ \bm{x}(t_{k})\in\calS_{T} $ as assumed, we obtain from~\eqref{eq:MPC thm 1} $ H_{T}(\bm{x}(t_{k+1})) \geq \min \lbrace H_{T}(\bm{x}(t_{k})), \delta \rbrace \geq 0 $. 
	Hence, $ \bm{x}(t)\in\calS_{T} $ for all $ t\in[t_{k},t_{k+1}] $.
	
	As it holds $ \bm{x}(0)\in\calS_{T} $ for $ t_{0}=0 $, we have by induction that $ \bm{x}(t)\in\calS_{T}\subseteq \calH $ for all $ t\geq 0 $. In particular, as $ \bm{x}(t_{k})\in\calH $ for all $ t_{k}\geq 0 $, it also follows that~\eqref{eq:MPC1} is recursively feasible which concludes the proof.
\end{proof}
\begin{remark}
	The MPC formulation~\eqref{eq:MPC1} is in so far limiting as the input maximizing~\eqref{seq:MPC max min} is directly applied on a time interval $ [0,\Delta t] $. In contrast,~\eqref{eq:safety filter} allows for any input that satisfies its CBF-based gradient constraint~\eqref{seq:safety filter constraint}. On the other hand, an MPC-formulation avoids the computation of a CBF. For answering the question if a CBF or an MPC approach is more suitable to ensure set-invariance, both the dimensionality of the system under consideration and the online control frequency are decisive. For lower dimesional systems whose control inputs shall be computed at a high frequency, a CBF-based approach is a good choice. In other cases, also an MPC approach can be considered.
\end{remark}
\begin{remark}
	As before, if Ass.~\ref{ass:control invariance} holds, then constraint~\eqref{seq:MPC terminal constraint} can be replaced by the terminal constraint $ \bm{x}(T;t_{k})\in\calF $.
\end{remark}

Flexibility can be added to the MPC-formulation by allowing for more general cost functions. Then, we can compute a control input that renders $ \calS_{T} $ invariant as
\begin{subequations}
	\label{eq:MPC2}
	\begin{align}
		\label{seq:MPC2 max min}
		&\hspace{-0.9cm}\min_{\bm{u}(\cdot;t_{k})} F(\bm{x}(\cdot;t_{k}),\bm{u}(\cdot;t_{k})) \\
		\label{seq:MPC2 dynamics}
		\text{s.t.}\;\;&\text{\eqref{seq:MPC dynamics}-\eqref{seq:MPC input constraint} hold,}\\
		\label{seq:MPC2 state constraint}
		&h(\bm{x}(s;t_{k})) \geq 0, \qquad \forall s\in[0,T] \\
		\label{seq:MPC2 terminal constraint}
		& \bm{x}(T;t_{k})\in\calF,
	\end{align}
\end{subequations}
where $ \bm{u}(\cdot;t_{k}): [0,T]\rightarrow \bbR^{m} $ is optimized over the time horizon $ T $; $ F $ is some cost function. In comparison to~\eqref{eq:MPC1}, we added the state constraint \eqref{seq:MPC2 state constraint} as the invariance of~$ \calS_{T} $ is not anymore necessarily implied by the cost function. 

\begin{theorem}
	\label{thm:MPC2}
	Let the same premises hold as in Cor.~\ref{thm:cbf simp}, and let $ \calS_{T} = \lbrace x \; | \; H_{T}(x)\geq 0 \rbrace $ where $ H_{T} $ is as defined in~\eqref{eq:H simp}. 
	If $ \bm{x}(0)\in\calS_{T} $, then an initially feasible solution to~\eqref{eq:MPC2} exists and~\eqref{eq:MPC2} is recursively feasible for $ t_{k}\geq0 $. Moreover, $ \bm{x}(t) \in \calS_{T} $ for all $ t\geq 0 $.
\end{theorem}
\begin{proof}
	Initial feasibility follows as before. Next, we assume that~\eqref{eq:MPC2} is feasible at time $ t_{k} $ and we denote the optimal input trajectory by $ \bm{u}^{\ast}(\cdot;t_{k}) $ and the corresponding optimal state trajectory by $ \bm{x}^{\ast}(\cdot;t_{k}):=\bm{\varphi}(\cdot;\bm{x}(t_{k}),\bm{u}^{\ast}(\cdot;t_{k})) $. Now, we consider the following candidate input trajectory~$ \bm{u}^{\text{c}} $ and its corresponding state trajectory~$ \bm{x}^{\text{c}} $ given as
	\begin{align*}
		\bm{u}^{\text{c}}(t;t_{k+1}) &:= \begin{cases}
			\bm{u}^{\ast}(t+\Delta t;t_{k}) & \text{if } t\in[0,T-\Delta t] \\
			\bm{u}_{e}(t) & \text{if } t\in(T-\Delta t,T]
		\end{cases} \\
		\bm{x}^{\text{c}}(t;t_{k+1}) &:= \begin{cases}
			\bm{x}^{\ast}(t+\Delta t;t_{k}) & \text{if } t\in[0,T-\Delta t] \\
			\bm{\varphi}(t;\bm{x}^{\ast}(T;t_{k}),\bm{u}_{e}(t)) & \text{if } t\in(T-\Delta t,T]
		\end{cases}
	\end{align*}
	where $ \bm{u}_{e}\in\bm{\calU}_{(T-\Delta t,T]} $ such that $ \bm{\varphi}(t;\bm{x}^{\ast}(T;t_{k}),\bm{u}_{e}(t))\in\calF $ for all $ t\geq T-\Delta t $. Such a control input exists because $ \bm{x}^{\ast}(T;t_{k})\in\calF $ and $ \calF $ is control-invariant according to Ass.~\ref{ass:control invariance}. Hence, $ \bm{x}^{\text{c}}(t;t_{k+1})\in\calF $ for all $ t\in(T-\Delta t,T] $. It follows now directly that constraints~(\ref{eq:MPC2}b-f) are satisfied by candidate trajectories $ \bm{u}^{\text{c}}(\cdot;t_{k+1}) $ and $ \bm{x}^{\text{c}}(\cdot;t_{k+1}) $ at time $ t_{k+1} $. Thereby, recursive feasibility is established. Furthermore as $ h(\bm{x}^{\ast}(t;t_{k}))\geq 0 $ and $ h(\bm{x}^{\text{c}}(t;t_{k+1}))\geq 0 $ for all $ t\in[0,T] $, it holds that $ H_{T}(\bm{x}(t_{k}+t'))\geq 0 $ for all $ t'\in[0,\Delta t] $ where $ \bm{x}(t_{k}+t') = \varphi(t';\bm{x}(t_{k}),\bm{u}^{\ast}(\cdot;t_{k})) $ as we consider nominal dynamics. By recursion, we conclude that $ \bm{x}(t)\in\calS_{T} $ for all~$ t\geq 0 $.
\end{proof}
In contrast to~\eqref{eq:MPC1}, we require a terminal constraint in~\eqref{eq:MPC2} in order to construct candidate trajectories that satisfy state constraint~\eqref{seq:MPC2 state constraint} and to establish recursive feasibility. Thereby, the additional flexibility in~\eqref{eq:MPC2} comes at the cost of the explicit knowledge on a control-invariant set. 
\begin{remark}
	The latter MPC formulation~\eqref{eq:MPC2} includes the predictive safety filter as proposed in~\cite{Hewing2020} as a special case. The idea of using predictions to extend a known control-invariant set has also been explored in~\cite{Gurriet2018a,Wabersich2023}. Although the knowledge on a control-invariant set also leads to simplifications in our case, it is not a requirement for our main results stated in Thm.~\ref{thm:cbf} and~\ref{thm:MPC recursive feasibility}.
\end{remark}


\section{Simulation}
\label{sec:simulation}



\begin{figure}[t]
	\centering
	\def\svgwidth{0.7\columnwidth}
	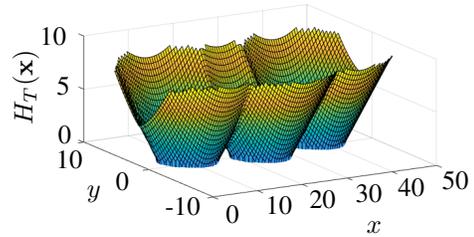
	\caption{CBF for scenario with three circular obstacles.}
	\label{fig:multi circ cbf}
	\vspace{0cm}
\end{figure}

\begin{figure}[t]
	\centering
	\def\svgwidth{0.7\columnwidth}
	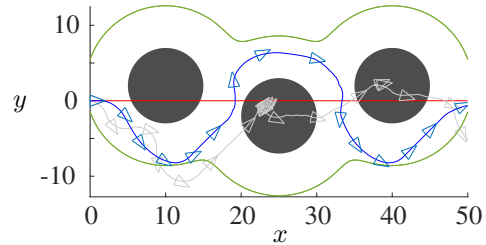
	\caption{Scenario with three circular obstacles.}
	\label{fig:multi circ sim}
	\vspace{-0.5cm}
\end{figure}

We consider again the vehicle with bicycle dynamics from Example~\ref{exmp:bicycle1}. Now, the vehicle uses a line following controller (nominal controller $ u_{\text{nom}} $) to follow a straight line. The safety filter~\eqref{eq:safety filter}, which is based on the CBF as computed in Example~\ref{exmp:bicycle4}, ensures that the vehicle avoids a circular obstacle. The simulation results are depicted in Fig.~\ref{fig:one circ sim}. 
The vehicle moves as closely as possible along the red line towards the right and avoids the circular obstacle. Its orientation is denoted by the triangles. The interior of the green circle marks those states for which the CBF has been computed. Similar results are obtained for a scenario with three circular obstacles, see Fig.~\ref{fig:multi circ cbf}-\ref{fig:multi circ sim}. However, if the computed CBF $ H_{T} $ is replaced by $ h $, the optimization problem~\eqref{eq:safety filter} may become infeasible and collision avoidance is not guaranteed anymore. The resulting trajectory is marked in Fig.~\ref{fig:multi circ sim} in light gray.

\section{Conclusion}
\label{sec:conclusion}

Based on mild controllability assumptions and a known subset of a possibly unknown control-invariant set, we showed how CBFs can be constructed by using predictions with a finite horizon. In doing so, we connected the construction of CBFs with the controllability properties of the system under consideration. With a relevant example, we outlined how the made assumptions can be satisfied. In the end, we related the prediction-based computation of CBFs to MPC.


\bibliographystyle{IEEEtran}
\bibliography{IEEEabrv,/Users/wiltz/CloudStation/JabBib/Research/000_MyLibrary}

\begin{thebibliography}{10}
\providecommand{\url}[1]{#1}
\csname url@samestyle\endcsname
\providecommand{\newblock}{\relax}
\providecommand{\bibinfo}[2]{#2}
\providecommand{\BIBentrySTDinterwordspacing}{\spaceskip=0pt\relax}
\providecommand{\BIBentryALTinterwordstretchfactor}{4}
\providecommand{\BIBentryALTinterwordspacing}{\spaceskip=\fontdimen2\font plus
\BIBentryALTinterwordstretchfactor\fontdimen3\font minus
  \fontdimen4\font\relax}
\providecommand{\BIBforeignlanguage}[2]{{%
\expandafter\ifx\csname l@#1\endcsname\relax
\typeout{** WARNING: IEEEtran.bst: No hyphenation pattern has been}%
\typeout{** loaded for the language `#1'. Using the pattern for}%
\typeout{** the default language instead.}%
\else
\language=\csname l@#1\endcsname
\fi
#2}}
\providecommand{\BIBdecl}{\relax}
\BIBdecl

\bibitem{Wieland2007}
P.~Wieland and F.~Allgöwer, ``Constructive safety using control barrier
  functions,'' \emph{IFAC Proceedings Volumes}, vol.~40, no.~12, pp. 462--467,
  2007, 7th IFAC Symposium on Nonlinear Control Systems.

\bibitem{Wills2004}
A.~G. Wills and W.~P. Heath, ``Barrier function based model predictive
  control,'' \emph{Automatica}, vol.~40, no.~8, pp. 1415--1422, 2004.

\bibitem{Wright2005}
M.~H. Wright, ``\BIBforeignlanguage{eng}{The interior-point revolution in
  optimization: History, recent developments, and lasting consequences},''
  \emph{\BIBforeignlanguage{eng}{Bulletin (new series) of the American
  Mathematical Society}}, vol.~42, no.~1, pp. 39--56, 2005.

\bibitem{Ames2019}
A.~D. {Ames}, S.~{Coogan}, M.~{Egerstedt}, G.~{Notomista}, K.~{Sreenath}, and
  P.~{Tabuada}, ``Control barrier functions: Theory and applications,'' in
  \emph{18th European Control Conference (ECC)}, 2019, pp. 3420--3431.

\bibitem{BlanchiniFranco2015SMiC}
F.~Blanchini and S.~Miani, \emph{\BIBforeignlanguage{eng}{Set-Theoretic Methods
  in Control}}.\hskip 1em plus 0.5em minus 0.4em\relax Springer International
  Publishing AG, 2015.

\bibitem{Artstein1983}
Z.~Artstein, ``\BIBforeignlanguage{eng}{Stabilization with relaxed controls},''
  \emph{\BIBforeignlanguage{eng}{Nonlinear analysis}}, vol.~7, no.~11, pp.
  1163--1173, 1983.

\bibitem{Sontag1989}
E.~D. Sontag, ``\BIBforeignlanguage{eng}{A ‘universal’ construction of
  artstein's theorem on nonlinear stabilization},''
  \emph{\BIBforeignlanguage{eng}{Systems \& control letters}}, vol.~13, no.~2,
  pp. 117--123, 1989.

\bibitem{Clark2021}
A.~Clark, ``Verification and synthesis of control barrier functions,'' in
  \emph{2021 60th IEEE Conference on Decision and Control (CDC)}, 2021, pp.
  6105--6112.

\bibitem{Wang2022}
\BIBentryALTinterwordspacing
H.~Wang, K.~Margellos, and A.~Papachristodoulou, ``Safety verification and
  controller synthesis for systems with input constraints,'' arXiv, 2022.
  [Online]. Available: \url{https://arxiv.org/abs/2204.09386}
\BIBentrySTDinterwordspacing

\bibitem{Xu2018a}
X.~Xu, J.~W. Grizzle, P.~Tabuada, and A.~D. Ames, ``Correctness guarantees for
  the composition of lane keeping and adaptive cruise control,'' \emph{IEEE
  Transactions on Automation Science and Engineering}, vol.~15, no.~3, pp.
  1216--1229, 2018.

\bibitem{Charitidou2021}
M.~Charitidou and D.~V. Dimarogonas, ``Barrier function-based model predictive
  control under signal temporal logic specifications,'' in \emph{19th European
  Control Conference (ECC)}, 2021, pp. 734--739.

\bibitem{Wabersich2023}
K.~P. Wabersich and M.~N. Zeilinger, ``Predictive control barrier functions:
  Enhanced safety mechanisms for learning-based control,'' \emph{IEEE
  Transactions on Automatic Control}, vol.~68, no.~5, pp. 2638--2651, 2023.

\bibitem{Gurriet2018a}
T.~Gurriet, M.~Mote, A.~D. Ames, and E.~Feron, ``An online approach to active
  set invariance,'' in \emph{2018 IEEE Conference on Decision and Control
  (CDC)}, 2018, pp. 3592--3599.

\bibitem{Squires2018}
E.~Squires, P.~Pierpaoli, and M.~Egerstedt, ``Constructive barrier certificates
  with applications to fixed-wing aircraft collision avoidance,'' in \emph{2018
  IEEE Conference on Control Technology and Applications (CCTA)}, 2018, pp.
  1656--1661.

\bibitem{Breeden2021}
J.~Breeden and D.~Panagou, ``High relative degree control barrier functions
  under input constraints,'' in \emph{2021 60th IEEE Conference on Decision and
  Control (CDC)}, 2021, pp. 6119--6124.

\bibitem{Breeden2022}
------, ``Predictive control barrier functions for online safety critical
  control,'' in \emph{2022 IEEE 61st Conference on Decision and Control (CDC)},
  2022, pp. 924--931.

\bibitem{Choi2021}
J.~J. Choi, D.~Lee, K.~Sreenath, C.~J. Tomlin, and S.~L. Herbert, ``Robust
  control barrier–value functions for safety-critical control,'' in
  \emph{2021 60th IEEE Conference on Decision and Control (CDC)}, 2021, pp.
  6814--6821.

\bibitem{Filippov1988}
A.~F. Filippov, \emph{\BIBforeignlanguage{eng}{Differential equations with
  discontinuous righthand sides}}, ser. Mathematics and its applications Soviet
  series, 18.\hskip 1em plus 0.5em minus 0.4em\relax Dordrecht: Kluwer Academic
  Publishers, 1988.

\bibitem{Khalil2002}
H.~K. Khalil, \emph{\BIBforeignlanguage{eng}{Nonlinear systems}}, 3rd~ed.\hskip
  1em plus 0.5em minus 0.4em\relax Prentice Hall, 2002.

\bibitem{Hermann1977}
R.~Hermann and A.~Krener, ``Nonlinear controllability and observability,''
  \emph{IEEE Transactions on Automatic Control}, vol.~22, no.~5, pp. 728--740,
  1977.

\bibitem{Wang2001}
D.~Wang and F.~Qi, ``Trajectory planning for a four-wheel-steering vehicle,''
  in \emph{Proceedings 2001 ICRA. IEEE International Conference on Robotics and
  Automation}, vol.~4, 2001, pp. 3320--3325.

\bibitem{Sussmann1987}
H.~J. Sussmann, ``A general theorem on local controllability,'' \emph{SIAM
  Journal on Control and Optimization}, vol.~25, no.~1, pp. 158--194, 1987.

\bibitem{Andersson2019}
J.~A.~E. Andersson, J.~Gillis, G.~Horn, J.~B. Rawlings, and M.~Diehl,
  ``{CasADi} -- {A} software framework for nonlinear optimization and optimal
  control,'' \emph{Mathematical Programming Computation}, vol.~11, no.~1, pp.
  1--36, 2019.

\bibitem{Hewing2020}
L.~Hewing, K.~P. Wabersich, M.~Menner, and M.~N. Zeilinger, ``Learning-based
  model predictive control: Toward safe learning in control,'' \emph{Annu. Rev.
  Control Robot. Auton. Syst.}, vol.~3, no.~1, pp. 269--296, May 2020.

\end{thebibliography}
\balance


\end{document}